\documentclass[aps,pra,twocolumn,showpacs,groupedaddress,letterpaper]{revtex4-1}

\usepackage{graphicx}
\usepackage{subfigure}
\usepackage{color}
\usepackage{physymb}
\usepackage{enumitem}
\usepackage{mdframed}

\begin{document}


\title{ACER: An Analytic Framework for Students' Use of Mathematics\\ in Upper-Division Physics}

\author{Bethany R. Wilcox}
\affiliation{Department of Physics, University of Colorado, Boulder, Colorado 80309, USA}
\author{Marcos D. Caballero}
\altaffiliation{Current address: Department of Physics and Astronomy, Michigan State University, East Lansing, Michigan 48824, USA}
\affiliation{Department of Physics, University of Colorado, Boulder, Colorado 80309, USA}
\author{Daniel A. Rehn}
\affiliation{Department of Physics, University of Colorado, Boulder, Colorado 80309, USA}
\author{Steven J. Pollock}
\affiliation{Department of Physics, University of Colorado, Boulder, Colorado 80309, USA}

\date{\today}

\begin{abstract}
Many students in upper-division physics courses struggle with the mathematically sophisticated tools and techniques that are required for advanced physics content.  We have developed an analytical framework to assist instructors and researchers in characterizing students' difficulties with specific mathematical tools when solving the long and complex problems that are characteristic of upper-division.  In this paper, we present this framework, including its motivation and development.  We also describe an application of the framework to investigations of student difficulties with direct integration in electricity and magnetism (i.e., Coulomb's Law) and approximation methods in classical mechanics (i.e., Taylor series).  These investigations provide examples of the types of difficulties encountered by advanced physics students, as well as the utility of the framework for both researchers and instructors.  
\end{abstract}


\maketitle

\section{\label{sec:Intro}Introduction}

Previous research has identified a considerable number of students' conceptual and mathematical difficulties, particularly at the introductory level (see Ref.\ \cite{Docktor2010} for a review).  Substantial work has also been done to characterize student problem solving in introductory physics \cite{Hsu2004}.  In addition to the significant work at the introductory level, researchers have recently begun to characterize students' conceptual knowledge in more advanced physics courses \cite{meltzer2012resource, Pepper2010, Wallace2010, Singh2006, Smith2010, Deslauriers2011, Ayene2011, Zhu2012}.  Furthermore, a small but growing body of research suggests that upper-division students continue to struggle to make sense of the mathematics necessary to solve problems in physics \cite{Pepper2012, Christensen2012, Sayre2008}. 

Upper-division physics content requires students to manipulate sophisticated mathematical tools (e.g. multivariable integration, approximation methods, special techniques for solving partial differential equations, etc.).  Students are taught these tools in their mathematics courses and use them to solve numerous abstract mathematical exercises.  Yet, many students still struggle to apply mathematical tools to problems in physics. This is not necessarily surprising given that physicists use mathematics quite differently than mathematicians (i.e., to make inferences about physical systems) \cite{Redish2005, Wagner2011}. However, persistent mathematical difficulties can undermine attempts to build on prior knowledge as our majors advance through the curriculum.  Upper-division instructors face significant pressure to cover new content, a task made more difficult by constantly having to review the relevant mathematical tools. It is often an explicit goal for advanced courses to develop students' ability to connect mathematical expressions to physics concepts. For example, consensus learning goals  for upper-division courses at the University of Colorado Boulder (CU) \cite{Pepper2011} include ``Students should be able to translate a physical description of an upper-division physics problem to a mathematical equation necessary to solve it,'' and ``to achieve physical insight through the mathematics of a problem.''  To improve student learning in advanced physics courses, we find it necessary to move away from merely noting students' conceptual difficulties towards systematically investigating how students integrate mathematics with their conceptual knowledge to solve complex physics problems.  

In order to address the issues that arise when solving physics problems that rely on sophisticated mathematical tools, we must first understand how students access and coordinate their mathematical and conceptual resources.  However, canonical problems in upper-division courses are often long and complex, and students' reasoning is similarly long and complex.   Making sense of the difficulties that arise requires a well-articulated framework for analyzing students' synthesis of conceptual knowledge and mathematical tools.  We use the term framework to refer to a structure of guiding principles and assumptions about the underlying relationship between a physical concept and the mathematics necessary to describe it.  At the upper-division level in particular, this relationship can be strongly dependent on the particular concept in question, suggesting that a useful framework needs to be adaptable to a wide variety of physical concepts and mathematical tools.  

We first encountered the need for such a framework while investigating students' understanding of approximation methods (i.e., Taylor series) in a middle-division classical mechanics course \cite{Caballero2012} and with integration of continuous charge distributions (i.e., Coulomb's Law) in an upper-division electrostatics course \cite{Wilcox2012}.  Our initial analysis focused on identifying emergent themes in students' work.  We quickly identified a multitude of common difficulties, but, beyond producing a laundry list of errors, we struggled to organize these issues in a productive way.  This lack of coherence made it challenging to identify relationships between the difficulties and to produce actionable implications for instruction or further research.  

To provide a suitable organizational structure, we developed a framework to address students' \emph{activation} of mathematical tools, \emph{construction} of mathematical models, \emph{execution} of the mathematics, and \emph{reflection} on the results (ACER).  The ACER framework is a tool designed to aid both instructors and researchers in exploring when and how students employ particular mathematical tools to solve canonical problems from upper-division physics courses.  Our goal is to provide a scaffold for describing student learning that is explicitly grounded in theories of learning but can still be leveraged by instructors who are not thoroughly versed in such theories.  

This paper serves the dual purpose of describing the theoretical grounding and development of the ACER framework (Sec.\ \ref{sec:Theory} and Sec.\ \ref{sec:ACER}) as well as presenting the methods and findings of two investigations of student difficulties at the upper-division level employing this framework (Sec.\ \ref{sec:App}).  It then closes with a discussion of limitations and implications for future work (Sec.\ \ref{sec:Discussion}).

\section{\label{sec:Theory}Problem-solving Strategies and Theoretical Frameworks}

There are two common aspects to understanding the problems students encounter when utilizing mathematics in physics.  The first is to characterize physicists' use of mathematics; such a characterization helps produce instructional and analytical tools to align students' problem solving with experts'.  The second is to describe what the students are actually doing, not just in terms of how it does not make sense to physicists, but in terms of how it does make sense to the students.  Here, we review some of the previous research using these two approaches.  

The first of these two aspects seeks to better understand the crossroads between physics and mathematics.  Redish \cite{Redish2005} has developed an idealized model of how physicists use math to describe physical systems.  He identifies four steps that guide this process: (1) map the physical structures to mathematical ones, (2) transform the initial mathematical structures, (3) interpret the results in terms of the physical system, and (4) evaluate the validity of the results.  This iterative model makes it clear that the source of students' difficulties may not be as simple as not knowing the necessary mathematical formalisms.  While the intentionally broad nature of the model makes it widely applicable, we found it challenging to utilize it to identify concrete, actionable implications for the instructor or researcher dealing with mathematical difficulties in the physics classroom.  

It has been well documented that students do not approach physics problems in a manner consistent with Redish's model \cite{Redish2005}.  In fact, students often approach physics problems in a way that seems haphazard and inefficient to experts \cite{Reif1976}.  Some attempts have been made to address this at the introductory level by explicitly teaching students a problem-solving strategy that is more aligned with the expert approach.  Wright and Williams \cite{Wright1986} incorporated a problem-solving strategy into their introductory physics course that involved four steps: (1) What's happening?, (2) Isolate the unknown, (3) Substitute, and (4) Evaluation (WISE).  The WISE strategy was designed as a heuristic that physics students could use to become more efficient and accurate problem solvers.  

Similarly, Heller et al.\ \cite{Heller1992} developed a strategy to help their introductory students integrate the conceptual and procedural aspects of problem solving.  This strategy included 5 steps: (1) Visualize the problem, (2) Physics description, (3) Plan the solution, (4) Execute the plan, and (5) Check and evaluate.  Docktor \cite{Docktor2009} modified and extended this strategy to develop a validated physics problem-solving assessment rubric.  With the goal of providing consistent and reliable scores on problem-solving tasks, this rubric is scored based on five general processes: Useful Description, Physics Approach, Specific Application of Physics, Mathematical Procedures, and Logical Progression. Useful Description is the process of summarizing a problem statement by assigning symbols and/or sketching.  Physics Approach and Specific Application of Physics represent the process of selecting and linking the appropriate physics concepts to the specifics of the problem.  Mathematical Procedures refers to the mathematical operations needed to produce a solution, and Logical Progression looks at the focus and consistency of the overall solution. 

The strategies presented above suggest considerable agreement as to the general structure of expert problem solving as well as some indication that this structure can be used as a guide to assess student work at the introductory level.  The prescriptive nature of these problem-solving strategies lends itself well to the kinds of problems encountered in introductory physics.  However, upper-division problems are more complex and less likely to respond to a prescriptive approach.  Additionally, problem-solving strategies are intentionally independent of specific content so as to be generally applicable, and on their own offer limited insight into the nature of students' difficulties with specific mathematical tools.  

The other aspect of understanding the problems students encounter when utilizing mathematics in physics focuses on explaining why students solve problems in a particular way.  Tuminaro \cite{Tuminaro2004} used videotaped problem-solving sessions with introductory students to develop a theoretical framework describing students' use of mathematics in physics.  This model of student thinking blends three theoretical constructs: mathematical resources \cite{Hammer2000}, epistemic games \cite{Collins1993}, and frames \cite{Tannen1993}.  Mathematical resources are the abstract knowledge elements that are involved in mathematical thinking.  Tuminaro \cite{Tuminaro2004} includes in the category of mathematical resources: a student's intuitive mathematics knowledge and sense of physical mechanism, their understanding of mathematical symbolism, and the strategies they use to extract information from equations.  Epistemic games are coherent patterns of activities observed during problem solving.  Each game is characterized by different sequences of moves and types of resources used by the student.  The game that a student chooses to play is governed by the frame they are operating in, which is determined by their tacit expectations for what kind of activity they are engaged in.  

The framework presented by Tuminaro \cite{Tuminaro2004} was developed for introductory students and relies on students' explicit discussion of the details of their work.  Upper-division students, on the other hand, tend to work more quickly and externalize less of their specific steps.  To address this, Bing \cite{Bing2008} leveraged the theoretical constructs of mathematical resources and epistemic framing to analyze upper-level students' use of mathematics.  Epistemic framing is the students' unconscious answer to the question `What kind of activity is this?'  Bing argues that a student's framing can be identified by examining the types of justifications and proof that they offer to support their mathematical claims, rather than the specific `moves' they make.  

There are several limitations to the theoretical frameworks from Tuminaro \cite{Tuminaro2004} and Bing \cite{Bing2008}.  To understand student work in terms of epistemic games or epistemic framing, one must have data on the students' real-time reasoning.  This largely restricts the potential data sources to video and audio data, eliminating students' written work.  Additionally, effective application of either framework requires considerable familiarity with the underlying theoretical constructs in PER.  In practice this will prevent many instructors, particularly at the upper-division level, from productively utilizing the frameworks.  

Describing experts' use of mathematics and characterizing students' problem solving are complementary aspects of understanding mathematical difficulties in physics.  The ACER framework leverages ideas from both in order to target students' use of mathematics in upper-division courses.

\section{\label{sec:ACER}The ACER Framework}

ACER is an analytical framework designed to guide and structure investigations of students' difficulties with the sophisticated mathematical tools used in their physics classes. When solving upper-division physics problems, students often make multiple mistakes or take unnecessary steps which must then be tracked through the solution. This undermines attempts to pinpoint the fundamental difficulties that cause the students to struggle or to identify relationships between these difficulties.  The ACER framework provides an organizing structure that focuses on important nodes in students' solutions.  This removes some of the ``noise'' in students' work that can obscure what is going on.  This section provides a general overview of the framework and its development before demonstrating its application to specific mathematical tools.  

\subsection{\label{sec:Overview}Overview}

ACER was developed in conjunction with research into student learning of two topics in upper-division physics: Taylor series \cite{Caballero2012} and direct integration \cite{Wilcox2012}. Direct integration and Taylor series were selected because they are representative of the kinds of mathematical tools that upper-division physics students are expected to use.  Additionally, previous work from both math and physics education suggest that these two topics are challenging for students \cite{Champney2012, Smith2013, Khan2011, Hu2013, Thompson2008, Kung2013}.  The results of applying the framework to these specific topics will be discussed in detail in Sec.\ \ref{sec:App}; here, we present the general development and form of ACER.  The ACER framework, like the frameworks presented by Tuminaro \cite{Tuminaro2004} and Bing \cite{Bing2008}, is fundamentally cognitive and assumes a resource view on the nature of knowledge \cite{Hammer2000}.  

In order to better understand students' difficulties, we performed a modified version of task analysis \cite{Catrambone2011, Catrambone1998} on canonical problems relating to each topic.   Task analysis is a method used to uncover the tacit knowledge used by experts when solving complex problems.  Our modified use of task analysis is described in greater detail in Sec.\ \ref{sec:Oper}; however, the general process requires a content expert to work through the problem while documenting and reflecting on all elements of a complete solution.  These elements are then discussed with several other content experts to reach consensus that all important aspects of the solution have been identified.  After several iterations, we found that these various problem-specific elements could be organized into four components that appeared consistently in the solutions to a number of content-rich problems utilizing sophisticated mathematical tools.  These four components are: \emph{Activation of the tool}, \emph{Construction of the model}, \emph{Execution of the mathematics}, and \emph{Reflection on the result}. Each component is described in greater detail below. 

In order to solve the back-of-the-book or exam-type problems that ACER targets, one must determine which mathematical tool is appropriate (Activation) and construct a mathematical model by mapping the particular physical system onto appropriate mathematical tools (Construction).  Once the mathematical model is complete, there is often a series of mathematical steps that must be executed in order to reduce the solution into a form that can be readily interpreted (Execution).  This final solution must then be interpreted and checked to ensure that it is consistent with known or expected results (Reflection).  The four general components are emergent from experts' problem solving and are consistent with previous literature on problem-solving strategies (see Sec.\ \ref{sec:Theory}).  Though the framework suggests a certain logical flow, we are not suggesting that all experts or students solve problems in a clearly organized, linear fashion. 

\begin{figure}
\centering
\includegraphics[clip, trim=50mm 40mm 20mm 40mm, width=0.70\linewidth, angle=90]{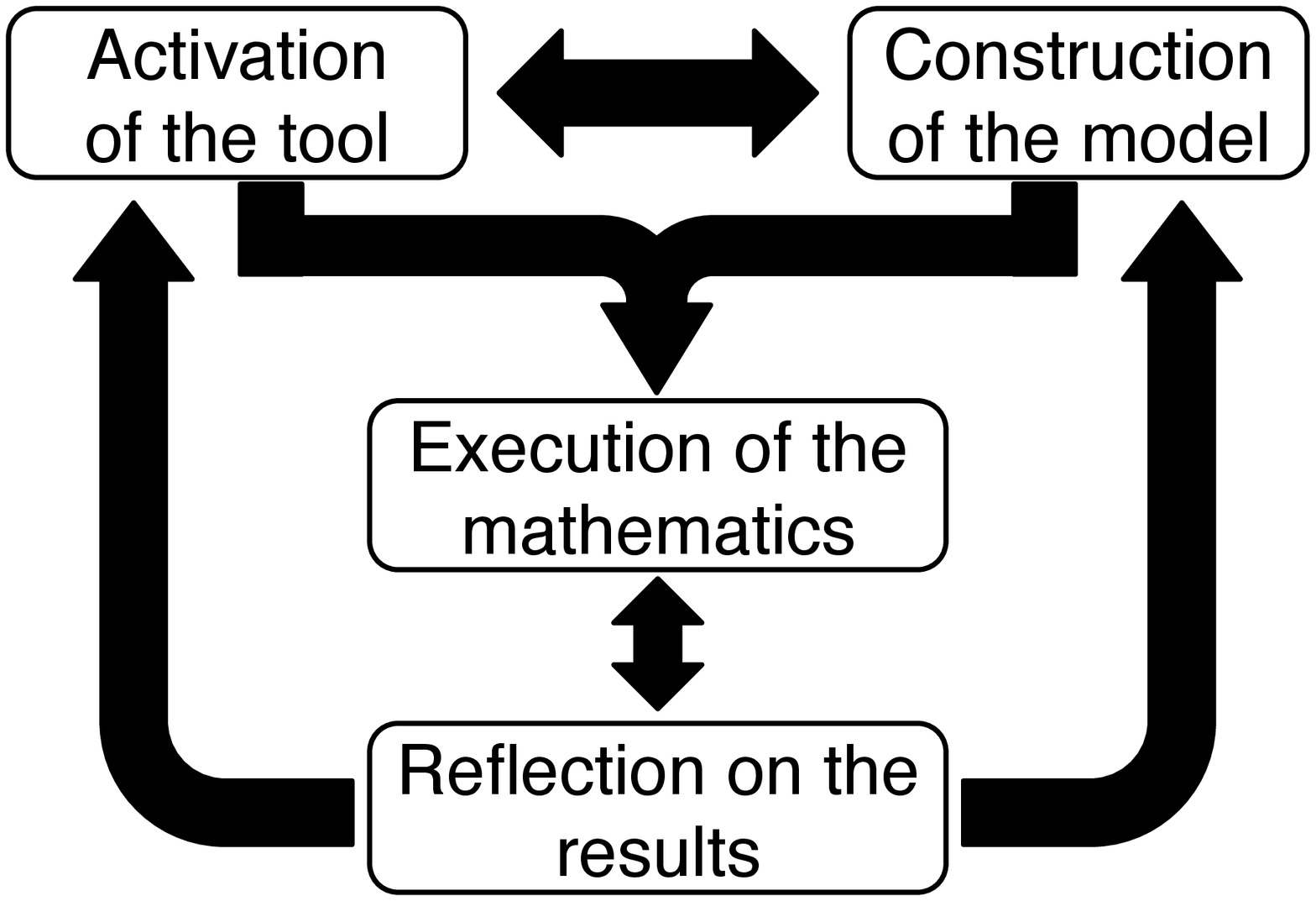}
\caption{A visual representation of the ACER framework.}\label{fig:acer}
\end{figure}

A convenient visualization of ACER is given in Fig.\ \ref{fig:acer}.  The framework provides a researcher-guided outline that organizes key elements of a well-articulated, complete solution.  The framework does not assign value by providing an ideal solution path towards which the students should strive.  ACER is also not designed to be general enough to be applied to open-ended problems; however, its targeted focus means it can be operationalized for a variety of mathematical tools used in context-rich problems.  Sec.\ \ref{sec:Oper} will provide several examples of how ACER is operationalized for specific tools and topics.   ACER is a tool for understanding and characterizing the difficulties seen in students' work, but its structure is not meant to approximate students' actual solutions.  Instead, the general structure of ACER was developed to accommodate the complex and often iterative solution patterns characteristic of upper-division problems.  

\textbf{Activation of the tool}: A problem statement contains a number of explicit and/or implicit cues that prime or activate different resources (or networks of resources) associated with any number of mathematical tools \cite{Hammer2000}.  These cues can include the goal of the problem (e.g., calculate the potential) as well as the language and symbols used.  The resources students activate depend on the individual student and their perception of the nature of the task (i.e., their epistemic framing \cite{Bing2008}).   

\textbf{Construction of the model}: In physics, mathematics are often used to express a simplified picture (i.e., a model) of a real system.  These mathematical models are typically necessary to solve physics problems.  Mathematical models are generally written in a remarkably compact form (e.g., $ \Delta\phi = - \int G\,dM/r$) where each symbol has a specific physical meaning, which may be context dependent.  Different representations (e.g., diagrammatic or graphical) are sometimes necessary to construct or map the elements of the model \cite{Redish2005}.

\textbf{Execution of the mathematics}: In order to arrive at a solution, it is usually necessary to transform the math structures produced in the construction component (e.g., unevaluated integrals) into mathematical expressions that can be more easily interpreted (e.g., evaluated integrals).  Each mathematical tool requires specific background knowledge and base mathematical skills (e.g., how to take derivatives or integrals).  The mathematical manipulations performed in this component are not necessarily context-free.  When employing these base mathematical skills, an expert maintains an awareness of the physical meaning of each symbol in the expression (e.g., which symbols are constants when taking derivatives or integrals) \cite{Redish2005}.  

\textbf{Reflection on the result}: Solutions to problems in upper-division physics usually result in expressions that are not merely superficial manipulations of formulas provided in the textbook or notes.  Instead, they are new entities that offer meaningful insight into explaining or predicting the behavior of physical systems.  Reflecting on these expressions is a crucial part of understanding the system and gaining confidence in the calculation performed (e.g., how do we know an expression is the correct one?).  At the most basic level, reflection involves checking expressions for errors (e.g., checking units) or comparing predictions to established or expected results (e.g., checking limiting behavior).  This kind of reflection can help to identify mistakes that occurred in the other components of the framework.  

The theoretical constructs that ground the frameworks presented by Tuminaro \cite{Tuminaro2004} and Bing \cite{Bing2008} are commensurate with the implicit theoretical constructs that ground ACER.  For example, a problem solver accesses different, possibly overlapping, networks of resources depending on the component of the framework in which they are working.  Similarly, certain epistemic frames would be more useful than others when operating in different components.  Bing identifies four epistemic frames used by upper-division students - Invoking Authority, Physical Mapping, Calculation, and Math Consistency \cite{Bing2008}.  Invoking Authority can be a valuable frame while in the Activation component. For instance, appealing to authority (e.g., the book or notes) is often a good way to identify which mathematical tool to use. In the Construction component, when trying to map that tool to a specific problem, relying on authority (e.g., depending on a similar problem in the book) can easily sidetrack the unwary student.  However, a Physical Mapping frame would likely be productive for both the Construction and Activation components.  While we acknowledge the value of leveraging theoretical constructs like resources and epistemic frames in conjunction with ACER, we have intentionally avoided explicit identification of specific resources or frames as part of the framework.  In this way, it is not necessary to have a strong background in theories of learning in order to utilize ACER.  

The general components of ACER were created by identifying broad themes that emerged from the modified task analysis of problems described in the next section.  These components are consistent with Redish's idealized model for the way physicists utilize mathematics \cite{Redish2005}, as well as the steps in the problem-solving strategies presented for introductory physics \cite{Reif1976, Wright1986, Heller1992, Docktor2009}.  Yet, ACER goes beyond these broad descriptions by providing a mechanism to target specific topics and mathematical tools.  This mechanism is described in the following section.  

\subsection{\label{sec:Oper}Operationalizing ACER}

The utility of ACER as a framework for understanding students' use of mathematics in physics comes when it is operationalized for a specific mathematical tool.  Operationalization is the process by which a particular problem or set of problems that exploit the targeted tool are mapped onto the framework.  This involves identifying important elements in each component that together result in what an expert/instructor would consider a complete and correct solution.  

We used a modified form of task analysis to operationalize the framework.  Formally, task analysis \cite{Catrambone2011, Catrambone1998} is accomplished by having a subject matter expert (SME) solve problems while explaining their steps and reasoning to a knowledge extraction expert (KEE) who keeps a record.  This method for uncovering the tacit knowledge used by experts has been exploited to produce example solutions designed to improve students' ability to solve novel problems \cite{Catrambone1996}.  

Our modified task analysis does not include a KEE.  This was done because such an expert was not readily available to us, nor did we want the need for a KEE to prevent other researchers or instructors from utilizing the framework.  Instead, the SME works through the problems, documenting their reasoning and mapping the vital elements of their solution onto the components of ACER.  This record is then shared with several other SMEs to ensure that all important aspects of the solution are accounted for.  Additionally, these experts come to a consensus in classifying each element into a specific component (i.e., Activation, Construction, Execution, or Reflection).  These preliminary elements are then applied to student work and the operationalized framework is refined to accommodate patterns of student reasoning not present in the SMEs solutions.  

Our motivation for removing the KEE was entirely practical in origin; however, not utilizing a KEE may have implications for the theoretical foundations of our modified task analysis.  The KEE, as a content novice, helps to force the SME to fully and clearly justify their steps even when they include decisions based on procedural and declarative details the SME no longer thinks about \cite{Catrambone2011}.  Removing the KEE from the task analysis process makes it more difficult to ensure that the important elements identified in the solution are complete from the point of view of a novice as well as an SME.  For this reason it is important that the operationalized ACER framework which is produced by the modified task analysis remains flexible to modification based on emergent analysis of student work.  

The following sections provide two examples of the operationalized framework from upper-division electrostatics and middle-division classical mechanics.  

\subsubsection{\label{sec:Coulomb}An Example from Electrostatics}

Determining the electric potential or electric field from a continuous charge distribution using the integral form of Coulomb's law is one of the first topics that upper-division students encounter in junior-level electrostatics.  For the remainder of the paper, we use Coulomb's Law to refer to the integral equation allowing for direct calculation of the electric field or potential from a continuous charge distribution.  

\begin{equation} 
E(\vec{r}) = \frac{1}{4 \pi \epsilon_0}\int_V \frac{dq}{|\vec{\scriptr}|^2} \hat{\scriptr} \label{eqn:Efield} 
\end{equation}
\begin{equation}
V(\vec{r}) = \frac{1}{4 \pi \epsilon_0}\int_V \frac{dq}{|\vec{\scriptr}|} \label{eqn:potential}
\end{equation}
Here, $dq$ represents the differential charge element and $\vec{\scriptr}$ is the difference vector $\vec{r}-\vec{r}'$ between the source and the observation location (i.e., Griffiths' \emph{script-r}) \cite{Griffiths1999}.  In this case, the `tool' we refer to is integration, and we describe its application to problems determining the potential or electric field from an arbitrary, static charge distribution via Coulomb's Law.  We will focus here only on charge distributions that cannot easily be dealt with using Gauss's Law.  The element codes  below are for labeling purposes only and are not mean to suggest a particular order nor are all elements always involved for any given problem.  

\textbf{Activation of the tool}: The first component of the framework involves the selection of a solution method.  The modified task analysis identified four elements that are involved in the activation of resources identifying direct integration (i.e., Coulomb's Law) as the appropriate tool. 

\vspace*{5pt}
 \begin{tabular}{lp{0.70\linewidth}}
   {\bf CA1} & The problem asks for the potential or electric field.\\ 
   {\bf CA2} & The problem gives a charge distribution.\\
   {\bf CA3} & The charge distribution does not have appropriate symmetry to productively use Gauss's Law.\\
   {\bf CA4} & Direct calculation of the potential is more efficient than starting with the electric field.
 \end{tabular}
\vspace*{5pt}

Elements {\bf CA1--CA3} are cues typically present in the problem statement. Element {\bf CA4} is specific to problems asking for the electric potential and is included to account for the possibility of solving for potential by first calculating the electric field.  This method is valid but often more difficult. 

\textbf{Construction of the model}: Here, mathematical resources are used to map the specific physical situation onto the general mathematical expression for Coulomb's Law.  The resulting integral expression should be in a form that could, in principle, be solved with no knowledge of the physics of this specific problem.  We identify four key elements that must be completed in this mapping.

\vspace*{5pt}
 \begin{tabular}{lp{0.70\linewidth}}
   {\bf CC1} & Use the geometry of the charge distribution to select a coordinate system. \\
   {\bf CC2} & Express the differential charge element ($dq$) in the selected coordinates. \\
   {\bf CC3} & Select integration limits consistent with the differential charge element and the extent of the physical system. \\
   {\bf CC4} & Express the difference vector, $\vec{\scriptr}$, in the selected coordinates. \\
  \end{tabular}
\vspace*{5pt}

Elements {\bf CC2} and {\bf CC4} can be accomplished in multiple ways, often involving several smaller steps.  In order to express the differential charge element, the student must combine the charge density and differential to produce an expression with the dimensions of charge (e.g., $dq = \sigma dA$).  Construction of the difference vector often includes a diagram that identifies vectors to the source point, $\vec{r}'$, and field point, $\vec{r}$.

\textbf{Execution of the Mathematics}: This component of the framework deals with the mathematics required to compute a final expression.  In order to produce a formula describing the potential or electric field, it is necessary to:

\vspace*{5pt}
 \begin{tabular}{lp{0.70\linewidth}}
   {\bf CE1} & Maintain an awareness of which variables are being integrated over (e.g., $r'$ vs. $r$). \\
   {\bf CE2} & Execute (multivariable) integrals in the selected coordinate system. \\
   {\bf CE3} & Manipulate the resulting algebraic expressions into a form that can be readily interpreted. \\
 \end{tabular}
\vspace*{5pt}       

\textbf{Reflection on the result}: The final component of the framework involves verifying that the expression is consistent with expectations.  While many different techniques can be used to reflect on the result, these two checks are particularly common: 

\vspace*{5pt}
 \begin{tabular}{lp{0.70\linewidth}}
   {\bf CR1} & Verify that the units are correct. \\
   {\bf CR2} & Check the limiting behavior to ensure it is consistent with the total charge and geometry of the charge distribution. \\
 \end{tabular}
\vspace*{5pt}

Element {\bf CR2} is especially useful when the student already has some intuition for how the potential or electric field should behave in the limits.  However, if they do not come in with this intuition, reflection on the results of this type of problem is a vital part of developing it.  

In Sec.\ \ref{sec:CoulombApp}, we will apply this operationalization of ACER to investigate student work on a canonical electrostatics problem (Fig.\ \ref{fig:Coulomb}).

\subsubsection{\label{sec:Taylor}An Example from Classical Mechanics}

Using Taylor series to construct an analytically-tractable problem, to approximate a complex expression, or to develop insight into a newly constructed solution are ubiquitous practices in physics.  
At CU, physics students typically first encounter Taylor series from a formal, mathematical perspective as freshman in calculus and then again as sophomores in their middle-division classical mechanics course from an applied physics perspective. 
We use Taylor series to refer to the general series approximation of continuous functions.

\begin{eqnarray}
f(x) &=& \sum_{n=0}^{\infty} \frac{1}{n!}f^{(n)}(x_0)(x-x_0)^n \nonumber \\
     &=& f(x_0) + f^{\prime}(x_0)(x-x_0) \nonumber \\
     & & \:\:\:\:\:\:\:\:\:\:\:\: + \frac{1}{2}f^{\prime\prime}(x_0)(x-x_0)^2 + \dots
     \label{eqn:TS}
\end{eqnarray}

Here, $f(x)$ represents some continuous function with continuous derivatives over the domain of interest. 
We will refer to $x-x_0$ as the {\it expansion parameter}, to $x$ as the {\it expansion variable}, and to $x_0$ as the {\it expansion point}. 
In this case, the ``tool'' we refer to is Taylor series, and its use is to describe approximations to complex expressions in order to gain insight about the underlying physics. 
In this paper, we will focus only on examples from classical mechanics though the framework could be applied to Taylor series in any domain.

\textbf{Activation of the tool}: 
The first component of ACER involves selecting Taylor series as an appropriate tool for a given problem. 
Our modified task analysis identified three elements that are likely to activate resources (or a network of resources) associated with Taylor series.

\vspace*{5pt}
 \begin{tabular}{lp{0.70\linewidth}}
   {\bf TA1} & The problem asks for a Taylor approximation directly. \\
   {\bf TA2} & The problem asks for an approximate expression to a complex function. \\
   {\bf TA3} & The problem uses language and/or symbols that imply one physical quantity is much smaller than some other physical quantity (e.g., ``small'', ``near'', ``close'', or $\ll$). \\
 \end{tabular}
\vspace*{5pt}

We include {\bf TA1} because Taylor series are often referred to explicitly in middle-division classical mechanics problems.
The physical quantities that are compared in {\bf TA3} must have the same units, and the ratio of these quantities must be less than 1.

\textbf{Construction of the model}: 
In this component, mathematical resources are used to map particular physical quantities onto the general expression for Taylor series (Eqn.\ \ref{eqn:TS}).  
After the mapping is complete, the approximation could in principle be completed with no additional knowledge of the physics of the problem.  For Taylor series, we identify four key elements to complete this mapping.

\vspace*{5pt}
 \begin{tabular}{lp{0.70\linewidth}}
   {\bf TC1} & Identify the physical quantities for which the problem explicitly states or implicitly suggests a comparison of scale (e.g., length scale, mass scale, time scale). \\
   {\bf TC2} & Determine about which point the comparison is being made (i.e., expansion point). \\
   {\bf TC3} & Express the comparison explicitly by constructing a dimensionless ratio of physical quantities (i.e., expansion variable). \\
   {\bf TC4} & Recast the expression to be expanded in terms of the expansion variable. \\
 \end{tabular}
\vspace*{5pt}

Element {\bf TC2} is often neglected because many approximations in physics are computed about zero (i.e., a Maclaurin series).  
In most problems, there are several combinations of physical quantities that could be used to construct a dimensionless ratio, but one must identify only those for which a comparison of scale is implied.  
Determining the appropriate expansion variable can be aided by sketching the physical situation and identifying the relative scales of physical quantities in the problem.

\textbf{Execution of the Mathematics}: 
This component of the framework is concerned with employing mathematics to compute a possible solution.  
Once the appropriate model has been constructed, the expansion can be computed.  
Strictly speaking, executing a Taylor series requires one to:

\vspace*{5pt}
 \begin{tabular}{lp{0.70\linewidth}}
   {\bf TE1} & Maintain an awareness of the meaning of each symbol in the expression (e.g., which symbols are constants when taking derivatives). \\
   {\bf TE2} & Compute derivatives of functions. \\
   {\bf TE3} & Evaluate the derivatives of non-trivial functions at the expansion point. \\
   {\bf TE4} & Manipulate the resulting algebraic expressions into a form that can be readily interpreted. \\
 \end{tabular}
\vspace*{5pt}

Alternatively, one might neglect elements {\bf TE2} and {\bf TE3}, if one has knowledge of common ``expansion templates'' (e.g., $\sin x \approx x - x^3/3!$) and how to adapt these templates to the mathematical models developed previously. Hence, there are two pathways to execute a Taylor series: a formal method involving all elements and an abbreviated method that shortcuts {\bf TE2} and {\bf TE3}.  The abbreviated method itself includes substeps, the details of which are beyond the scope of this study and thus have not been articulated here.

\textbf{Reflection on the result}: 
The final component describes how to verify that the approximate expression is consistent with expectations.  The expressions that result from performing a Taylor series are often novel entities, not superficial manipulations of formula from textbooks or notes, and these expressions must be checked.

\vspace*{5pt}
 \begin{tabular}{lp{0.70\linewidth}}
   {\bf TR1} & Verify that the units are correct. \\
   {\bf TR2} & Check the behavior in the regime where the approximation applies to ensure it is consistent with prior knowledge or intuition about the physical system.  \\
 \end{tabular}
\vspace*{5pt}

This component is particularly important for Taylor series because such approximations are used to check or make sense of solutions to many other problems.

In Sec.\ \ref{sec:TaylorApp}, we will apply this operationalization of ACER to investigate student work on several Taylor series problems.

\section{\label{sec:App}Application of ACER}

To demonstrate the utility and versatility of ACER, we present findings from two investigations of student difficulties in the advanced physics courses at CU: direct integration of continuous charge distributions and Taylor series as an approximation method.  These investigations were conducted independently as part of broader transformation efforts associated with CU's upper-division Principles of Electricity and Magnetism 1 (E\&M 1) course \cite{chasteen2012thinking,chasteen2012transforming} and middle-division Classical Mechanics and Mathematical Methods 1 (CM 1) course \cite{Pollock2011}.  Data for these studies come from analysis of student solutions to traditional exam questions and formal, think-aloud interviews.  In both cases, initial data collection and analysis began prior to the development of the ACER framework.  Application of the framework to initial data motivated a second round of interviews for both topics.  This section presents the methods and findings of these two investigations with particular emphasis on how ACER contributed to the analysis.  

\subsection{\label{sec:Methods1}Background}

Data for these studies were collected in association with the E\&M 1 and CM 1 courses at CU. Below, we provide additional details on the methods for our direct integration of Coulomb's Law (Sec.\ \ref{sec:CLmethods}) and Taylor series (Sec.\ \ref{sec:TSmethods}) studies. E\&M 1 typically covers the first 6 chapters of Griffiths \cite{Griffiths1999}, which includes both electrostatics and magnetostatics.  CM 1 uses Boas \cite{Boas2006} along with Taylor \cite{Taylor2005} and covers up to but not including calculus of variations.  The student population for both courses is composed of physics, engineering physics, and astrophysics majors, with a typical class sizes of 30-70 students.  These courses have been transformed to include a number of research-based teaching practices including peer instruction \cite{Mazur1997} using clickers and tutorials \cite{chasteen2012thinking,chasteen2012transforming}.  

In order to determine the types of difficulties students have with Coulomb's Law integrals and Taylor series, we analyzed student solutions to canonical exam problems on continuous charge distributions (N=172) and approximation methods (N=116) and conducted two sets of think-aloud interviews (Total N=18) to further probe student understanding.  The specific details of each exam problem is described in greater detail below.  Interviews were videotaped and students' written work was captured with embedded audio.  Interviewees were paid volunteers who responded to an email request for research participants.  All interviewees had successfully completed E\&M 1 or CM 1 one to two semesters prior.  Participants in both studies demonstrated a wide range of abilities and received course scores ranging from A to D.  

Exams were analyzed by identifying each of the key elements from the framework that appeared in the students' solutions. Each element was then coded to identify the types of steps made by students.  These codes represented emergent themes in the students' work around each element and were not predetermined by the framework.  This coding helped to ensure that the expert-guided framework did not miss important but unanticipated aspects of student solutions.  The interviews were similarly analyzed by classifying each of the student's major moves into one of the four components of the framework.  Exams provided quantitative data identifying common difficulties and interviews offered deeper insight into the nature of those difficulties.  

\subsection{\label{sec:CoulombApp}Coulomb's Law}

\begin{figure}
\begin{mdframed}
\vspace{2mm}\flushleft Calculate the electric potential at point P on the $z$-axis from a disk with a given surface charge density $\sigma(\phi)$.
\begin{center}
\includegraphics[clip, trim=65mm 210mm 80mm 18mm, width=0.90\linewidth]{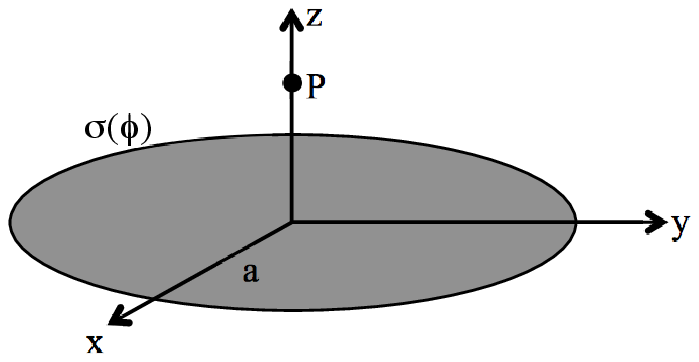}
\end{center}
\end{mdframed}
\caption{An example of the canonical exam problem on continuous charge distributions.} \label{fig:Coulomb}
\end{figure}

\vspace{-14pt}
\subsubsection{\label{sec:CLmethods}Methods}

Our E\&M 1 students are exposed to the Coulomb's Law integral for the electric field (Eqn.\ \ref{eqn:Efield}) before the analogous expression for the electric potential (Eqn.\ \ref{eqn:potential}).  However, the vector nature of the electric field makes Eqn.\ \ref{eqn:Efield} significantly more challenging to calculate, and historically, instructors at CU tend to ask students to compute the potential on exams.  The exam problem examined here asked students to calculate the electric potential along an axis of symmetry from a disk with charge density $\sigma(\phi)$ (Fig.\ \ref{fig:Coulomb}).  We selected this problem because it is a recognizable Coulomb's law question which requires integration and has been asked on the first midterm exam for multiple semesters. 

Exams were collected from four semesters of the course (N=172), each taught by a different instructor.  Two of these instructors were physics education researchers involved in developing the transformed materials and two were traditional research faculty. All four semesters utilized some or all of the available transformed materials. The exact details of the disk question, while similar, were not identical from semester to semester.  One of the PER faculty asked the students to sketch the charge distribution and then to calculate an expression for the potential on the z-axis (as in Fig.\ \ref{fig:Coulomb}).  The other PER faculty asked the students to calculate the total charge on the disk but only required them to set up the expression for the potential on the x-axis as the resulting integral cannot be solved easily by hand.  Both non-PER faculty asked for the total charge on the disk first and then for the potential on the z-axis.  

\begin{figure}
\begin{mdframed}
\includegraphics[clip, trim=52mm 191mm 68.5mm 44mm, width=1\linewidth]{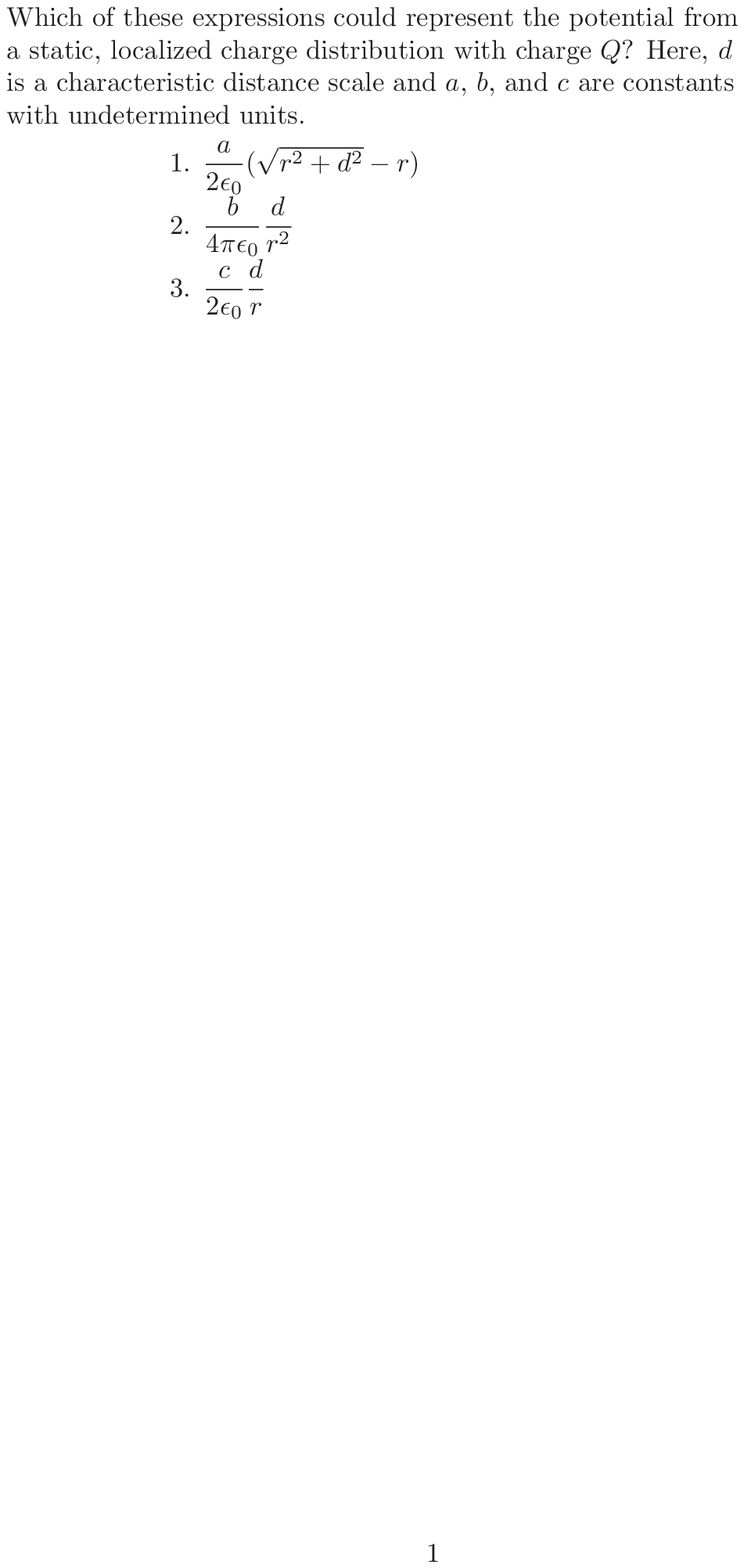}
\end{mdframed}
\caption{Three equations presented in the second interview set to target Reflection.  Students must determine the units of $a$, $b$, and $c$.} \label{fig:Reflection}
\end{figure}

Interview data came from two sets of think-aloud interviews (N=10), performed approximately 1 year apart on different sets of students.  The first set of interviews was structured to probe the preliminary difficulties identified in the student exams.  The students were asked to calculate the potential from two parallel disks of charge by direct integration, and they were provided with a diagram of the charge distribution and Eqn.\ \ref{eqn:Efield} and \ref{eqn:potential}.  In terms of the ACER framework, this prompt completely bypassed the Activation component.  Also, while the first interview protocol offered important insight into how students spontaneously reflect (or not) on their solutions, it provided no explicit probe of the Reflection component.  The second interview protocol specifically targeted Activation by asking students to find the potential along the z-axis outside a spherical shell with non-uniform charge density $\sigma(\theta)$ without providing a diagram or prompting them to solve the problem in any specific manner. An additional question targeted Reflection by asking students to determine which of three expressions could represent the potential from a static, localized charge distribution with total charge Q (see Fig.\ \ref{fig:Reflection}).

\subsubsection{\label{sec:Results1}Results}

This section presents the identification and analysis of common student difficulties with Coulomb's Law integrals organized by component and element of the operationalized ACER framework (See Sec.\ \ref{sec:Coulomb}).  

\textbf{Activation of the tool}: Roughly three-quarters of our students (73\% of 172) correctly approached the exam question using Eqn.\ \ref{eqn:potential}.  The remaining students (27\% of 172) attempted to calculate the potential by determining $\vec{E}$, either by Gauss's Law or Eqn.\ \ref{eqn:Efield}, and then taking the line integral (i.e., missing elements CA3 and CA4).  Rather than stemming primarily from a failure to recall Eqn.\ \ref{eqn:potential}, we argue below that this difficulty likely originated from a failure to reject these other methods.

Identifying evidence of Activation in the exam solutions was challenging because students did not typically write out their thought process as they began the problem.  In particular, there was rarely explicit evidence that the students attended specifically to CA1 and CA2 (i.e., the prompt asked for potential and provided information on the charge distribution).  However, we did not see students attempting to calculate quantities unrelated to the potential or attempting to utilize methods inconsistent with the information provided.  

More easily identified was element CA3, which eliminates Gauss's Law as a valid approach.  Approximately a tenth of our students (11\% of 172) attempted to employ Gauss' Law to solve for $\vec{E}$ and then to calculate V by taking line integral.  These students often justified their answers with comments such as, ``Since we want the voltage at a point outside the disk, the E-field we use will appear to be that of a point charge at the origin.''  This inappropriate use of Gauss's Law is consistent with previous research at the junior-level \cite{Pepper2012}. Interestingly, none of the students in the single semester (N=25) that were asked to sketch the charge distribution rather than to calculate total charge attempted to use Gauss's Law.  This suggests that calculation of the total charge likely activated resources associated with Gauss's Law.  

The misapplication of Gauss's Law was also the primary issue observed in the interviews.  Even when the students were explicitly prompted to use direct integration, one of five students still attempted to use Gauss's Law.  Two students in the second set of interviews explicitly considered using Coulomb's Law but rejected it in favor of using Gauss's Law or the expression for E from a point charge.  ACER states that there are a number of cues (elements CA1--CA3) embedded in the prompt of a physics problem that can guide a student to the appropriate solution method.  For example, if the prompt provides a boundary condition rather than a charge distribution, this is likely to cue the student to use separation of variables or method of images.  Elements CA1 and CA2 are identical for questions that can be solved by Gauss's Law and Coulomb's Law (i.e., it asks for V or E and provides $\rho(\vec{r}')$).  However, our students tend to be more comfortable with Gauss's Law (i.e., their Gauss's Law resources are easily activated); therefore, they must first reject Gauss's Law as appropriate before they will attempt to use Coulomb's Law.   

Even without Gauss's Law, it is still possible to solve for V by first calculating E using Eqn.\ \ref{eqn:Efield}, but this calculation requires considerably more work (element CA4).  Indeed, of the students who attempted this method (15\% of 172) only a few (N=3) completed the exam problem successfully.  One virtue of the electric potential in electrostatics is to allow for easier calculation of the electric field via $\vec{E}=-\vec{\nabla}V$.  However, the students may have jumped to calculating V from E because they were exposed to $\vec{E}$ first and resources associated with the electric field were more easily activated.  This difficulty was not observed in the interviews.  

\textbf{Construction of the model}: For Coulomb's Law integrals, the largest number of common student difficulties appeared in the Construction component, particularly when expressing the differential charge element and difference vector (elements CC2 and CC4).  These difficulties cannot be explained purely by students failing to conceptualize the integral or lacking the mathematical skills to set up integrals over surfaces and perform vector subtractions.  Rather, students had trouble keeping track of the relationships between various quantities as they adapted the deceptively simple general formula (Eqn.\ \ref{eqn:potential}) to a specific physical system.    

Almost all of the exams (97\% of 172, N=166) contained elements from the Construction component (i.e., the student did more than just write down the equation).  Of these students, only two did not use the appropriate coordinates (i.e., cylindrical), indicating that students at this level are adept at selecting appropriate coordinate systems in highly symmetric problems (element CC1).  Similarly, only one of the interview participants started with an inappropriate coordinate system, and this student eventually switched after attempting the problem in Cartesian coordinates.  This finding is somewhat surprising given prior research indicating that even middle-division physics students often have a strong preference for Cartesian coordinates \cite{Sayre2008}.  

The remaining elements of Construction proved more challenging.  Nearly half the students (42\% of 166) had difficulty expressing the differential charge element (element CC2) and some (14\% of 166) failed to provide limits of integration or gave limits that were inconsistent with their differential (element CC3).  The most common errors made while expressing the differential charge element ($dq$) were (see Table \ref{tab:dq}):  performing the integration over a region of space with zero charge density, using a differential with the wrong units, and plugging in total charge instead of charge density.

\begin{table}
\caption{Difficulties expressing the differential charge element ($dq$). Percentages are of just the students who had difficulty with $dq$ (42\% of 166, N=69).  Codes are not exhaustive or exclusive but represent the most common themes, thus the total N in the table need not sum to 69.}\label{tab:dq}
  \begin{ruledtabular}
    \begin{tabular} {l c c}
	\textbf{Difficulty} 				& \textbf{N} 	& \textbf{Percent} 					\\
	\hline
	Not integrating only over charges 	& 37 			& 54 							\\
	\hspace{3mm} e.g., $dq = \sigma \hspace{1mm} dr \hspace{1mm} dz \hspace{1mm} rd\phi$ & 	& 	\\
	Differential with the wrong units  	& 23 			& 33 							\\
	\hspace{3mm} e.g., $dq = \sigma \hspace{1mm} dr \hspace{1mm} d\phi$ 			& 	& 	\\
	Total charge instead of charge density  	& 10 			& 14 							\\
	\hspace{3mm} e.g., $dq = Q_{tot} \hspace{1mm} dr \hspace{1mm} rd\phi$	 		& 	& 	\\
    \end{tabular}
  \end{ruledtabular}
\end{table}

Initially, we interpreted difficulties with $dq$ as a failure to conceptualize Eqn.\ \ref{eqn:potential} as a sum over each little `bit' of charge. Previous research on student difficulties with the concept of accumulation as it applies to definite integrals supports this interpretation \cite{Thompson2008}.  However, the interviews suggest that the problem was more subtle than that.  Even those interviewees who failed to produce an appropriate expression for $dq$ made statements or gestures indicating they understood the integral to be a sum over the charge distribution.  Additionally, post-test data from the classical mechanics course at CU shows that more than 80\% of our students can correctly determine the differential area element for a cylindrical shell one semester prior to taking E\&M.  Thus the problem appeared to be neither that the students were not conceptualizing the integral as a sum over the charges, nor that they could not construct a differential area element.  Instead, the difficulties appeared when students were asked to apply these two ideas simultaneously to produce an expression for $dq$ consistent with a specific charge distribution.  

The magnitude of the difference vector, $|\vec{\scriptr}|$, must also be expressed such that it is consistent with the specific charge distribution (element CC4), and most students (86\% of 172, N=148) attempted to do so.  About half of these (47\% of 148) were unable to produce a correct formula for $|\vec{\scriptr}|$.  The most common errors included (see Table \ref{tab:scriptr}): using a magnitude appropriate for a ring of charge, setting the magnitude equal to the distance to the source point ($r'$), setting the magnitude equal to the distance to the field point ($r$), and never expressing the magnitude in terms of given variables or quantities. It was difficult to distinguish between the middle two difficulties because students' notation rarely distinguished clearly between the source and field variables; these issues are combined in Table \ref{tab:scriptr}. The remaining students were distributed over a variety of distinct, but not widely-represented issues.

\begin{table}
\caption{Difficulties expressing the magnitude of the difference vector ($\scriptr$). Percentages are of just the students who had difficulties with $\scriptr$ (47\% of 148, N=69). Codes are not exhaustive but represent the most common themes, thus the total N in the table need not sum to 69.}\label{tab:scriptr}
  \begin{ruledtabular}
    \begin{tabular} {l c c}
	\textbf{Difficulty} 						& \textbf{N} 	& \textbf{Percent} 	\\
 	\hline
	Ring of charge 						& 27 			& 39 			\\
	\hspace{3mm} i.e., $|\vec{\scriptr}| = \sqrt{a^2+r'^2}$ 	& 		& 			\\
	Distance to source or field point 				& 17 			& 25 			\\
	\hspace{3mm} i.e., $|\vec{\scriptr}| = r$ or $|\vec{\scriptr}| = r'$ & 	& 			\\
           No expression for $|\vec{\scriptr}|$			& 8			& 12			\\
    \end{tabular}
  \end{ruledtabular}
\end{table}

Students' spontaneous use of diagrammatic representation may be an additional aspect of the Construction component.  For example, drawing the vectors $\vec{r}$, $\vec{r}'$, and $\vec{\scriptr}$ is a helpful step towards a correct expression for $|\vec{\scriptr}|$.  We found that about two-thirds our students (66\% of 148, N=98) drew one or more of these vectors on the exams; however, only half of these students (50\% of 98) made explicit use of this diagram in their solution.  It may be that our students have seen enough of these types of problems to know that they should draw a diagram but have not internalized how to use it productively.   

Six of the eight interview participants who used Coulomb's Law also spontaneously drew the difference vector, and a seventh drew the vector but did not explicitly identify it as $\vec{\scriptr}$.  However, even those students who were able to articulate the difference vector as the distance between the source and field point struggled to produce a useful expression for it.  Only one interview participant arrived at a correct expression for the difference vector while the others were either unable to express $|\vec{\scriptr}|$ or treated it as a single variable like $r$ or $r'$.  The greater degree of difficulty with $\vec{\scriptr}$ observed in the interviews may be due to the time delay between the participants completing the course and sitting for the interview.  

Using Griffith's ``script-r'' notation, rather than $\vec{r}-\vec{r}'$, has a number of advantages including making Coulomb's law for continuous charge distributions look very similar to Coulomb's law for a point charge.  However, it may be that this notation also encourages students to look at $\vec{\scriptr}$ as a separate entity that they must remember rather than a quantity they construct.  In fact, most students made comments in the interviews about not remembering the formula for $\vec{\scriptr}$ or which direction it pointed, and few even attempted to use the source and field point vectors to answer these questions.  Only three of the eight interviewees spontaneously drew $\vec{r}$ and $\vec{r}'$, suggesting that the ``script-r'' notation obscured the importance of these two vectors. Failure to properly distinguish between $\scriptr$, $r$, and $r'$ often resulted in improper cancellations in the Execution component.  

\textbf{Execution of the mathematics}: Given the high pressure and individual nature of both exams and interviews, we expected that many students would make mathematical errors particularly with element CE3.  Yet our data offer no evidence that mathematical errors either with integrals or algebraic manipulations (elements CE2 or CE3) were specific to solving Coulomb's Law problems nor that they represented the primary barrier to student success on these problems.  More than half the student exam solutions (60\% of 172) contained elements from the Execution component.  The significant reduction in number was due primarily to the one of the four classes (N=55) that was only asked to set up the integral for V.  Additionally, not all students progressed far enough in their solutions to actually evaluate integrals.  

We were not able to produce a quantitative measure of student difficulties with element CE1 from the exams because the majority of students did not consistently distinguish between source and field variables (i.e., $r$ vs.\ $r'$).  However, of the four interview participants who made a distinction between the source and field point, none consistently used the primed notation. Two of these students ended up integrating over the $r$ variable as if it were $r'$.  

Overall, half the students' exams containing elements of Execution (51\% of 103, N=53) made various mathematical errors while solving integrals or simplifing their expression algebraically (elements CE2 and CE3).  Roughly half of the students with mathematical errors (49\% of 53) made only slight mathematical errors, such as dropping a factor of two or plugging in limits incorrectly.  The remaining students (51\% of 53) made various significant mathematical errors, such as pulling integration variables outside of integrals or not completing one or more integrals.  Similar trends were observed with the seven interview participants who attempted to complete one or more calculations.  Four students made significant mathematical errors, two made only slight mathematical errors, and one made no errors.  

\textbf{Reflection on the result}: In many cases, mistakes in the Construction or Execution component resulted in expressions for the potential which had the wrong units and/or limiting behavior (elements CR1 and CR2). While our students were able to identify these checks as valuable when explicitly prompted, we found that they rarely spontaneously check these properties to gain confidence in their solutions.  

Only a small number of students (8\% of 172) made explicit attempts to check their work on exams and almost exclusively by checking limiting behavior.  While it is possible that a greater number of students did perform one or more checks (i.e., elements CR1 and CR2) but simply did not write them out, the interviews suggest this is less likely.  When they had not been prompted to check or reflect on their solutions, half of the interview participants made no attempt to do so.  Two of the remaining students only made superficial comments about being uncertain if their solution was correct.  One stated that her answer did not makes sense but was not able to leverage this realization to correct her earlier work.  The final two students both mentioned checking the units of their solutions, though not recalling the units of $\epsilon_o$ prevented one of them from actually doing so.  

The second set of interviews explicitly targeted Reflection by directly asking the students to determine if three formulas (Fig.\ \ref{fig:Reflection}) could represent the potential from a static, localized charge distribution with positive total charge Q.  All five students suggested checking the units of these expressions, yet all but one had difficulty doing so because they did not recall the units of $\epsilon_o$.  This may be part of why units checks were not more common in the exam solutions as well.  Eventually, all the students were able to execute a units check once shown a method for getting around the units of $\epsilon_o$ by considering the formula for the potential of a point charge.  Additionally, all five students suggested checking that in the limit as $r \to \infty$ the potential went to zero.  Only two students spontaneously argued that V would need to fall off as $\frac{1}{r}$.  The other three made this argument when their attention was specifically drawn to the fact that the charge distributions was localized and had positive total charge.  

One of the three expressions for V required an appropriate Taylor expansion in order to determine its behavior at large $r$ (i.e., expression 1 of Fig.\ \ref{fig:Reflection}).  Only one of the five students recognized the need for an expansion without prompting.  Another three argued that the expression clearly did not fall off like a point charge.  However, when directed to Taylor expand, all three were able to manipulate the expression in order to isolate the small quantity and determine the leading term in the series.   A more detailed discussion of student difficulties with Taylor series through the lens of ACER is given in Sec.\ \ref{sec:TaylorApp}.  

\subsubsection{\label{sec:Imp1}Summary \& Implications}

We found that our junior-level students tended to encounter two broad difficulties which inhibited them from successfully solving for the potential from a continuous charge distribution using Coulomb's Law.  First, students struggled to activate direct integration via Coulomb's Law as the appropriate solution method.  In particular, some students tried to calculate the potential by first calculating the electric field by Gauss's Law or Coulomb's Law.  For instructors, this suggests that presentation of Eqn.\ \ref{eqn:potential} should be accompanied by explicit emphasis on \emph{when} and \emph{why} Gauss's Law cannot be used as well as the utility of calculating the electric potential rather than the electric field.  The latter should be aimed at helping students to develop strong connections between the conceptual idea of the potential and various mathematical formula which allow them to calculate V($\vec{r}$).  Second, students had difficulty coordinating their mathematical and physical resources to construct an integral expression for the potential which was consistent with the particular physical situation, specifically when expressing the differential charge element, $dq$, and difference vector, $\vec{\scriptr}$.  Instructors may be able to help by highlighting the relationships between these quantities to encourage students to view  Eqn.\ \ref{eqn:potential} a as coherent whole rather than a conglomeration of disconnected pieces.  We also found that while our juniors were capable of correct and meaningful reflection when explicitly prompted, very few executed these reflections spontaneously.  We consider the ability to translate between physical and mathematical descriptions of a problem and to meaningfully reflect on or interpret the results as two defining characteristics of a physicist, yet these are areas where our students struggled most when manipulating Coulomb's Law integrals.   

\subsection{\label{sec:TaylorApp}Taylor Series}

\begin{figure}        
        \subfigure[]{
        \begin{minipage}{\linewidth}
        \begin{mdframed}\vspace*{2mm}
                \flushleft The horizontal motion of a projectile experiencing linear drag is given by $x(t) = \dfrac{v_{x0}m}{b}\left(1-e^{-\frac{bt}{m}}\right)$.\\
                (i) For very small time, obtain an approximate expression for $x(t)$ with at least two non-zero terms using an appropriate Taylor expansion.\\
                (ii) What is the significance of each of the terms in your answer?\\
                (iii) For $t \approx m/b$, obtain an approximate expression for $x(t)$ using an appropriate Taylor expansion.\\\vspace*{2mm}
        \end{mdframed}
        \end{minipage}
        \label{fig:drag}}
        
        \subfigure[]{
		\begin{minipage}{\linewidth}
        \begin{mdframed}\vspace*{2mm}
                \flushleft A small sphere (mass, m) is free to slide inside a frictionless cylinder of radius $R$. If placed at the equilibrium point, $\phi = 0$ (shown below), the ball does not move. The gravitational potential energy for this system is given by $U(\phi) = A (1 - \cos \phi)$.\\\vspace*{1mm}
                \begin{minipage}{0.35 \linewidth}
				\begin{center}
					\includegraphics[width=\linewidth, clip, trim=80mm 40mm 30mm 80mm, angle=90]{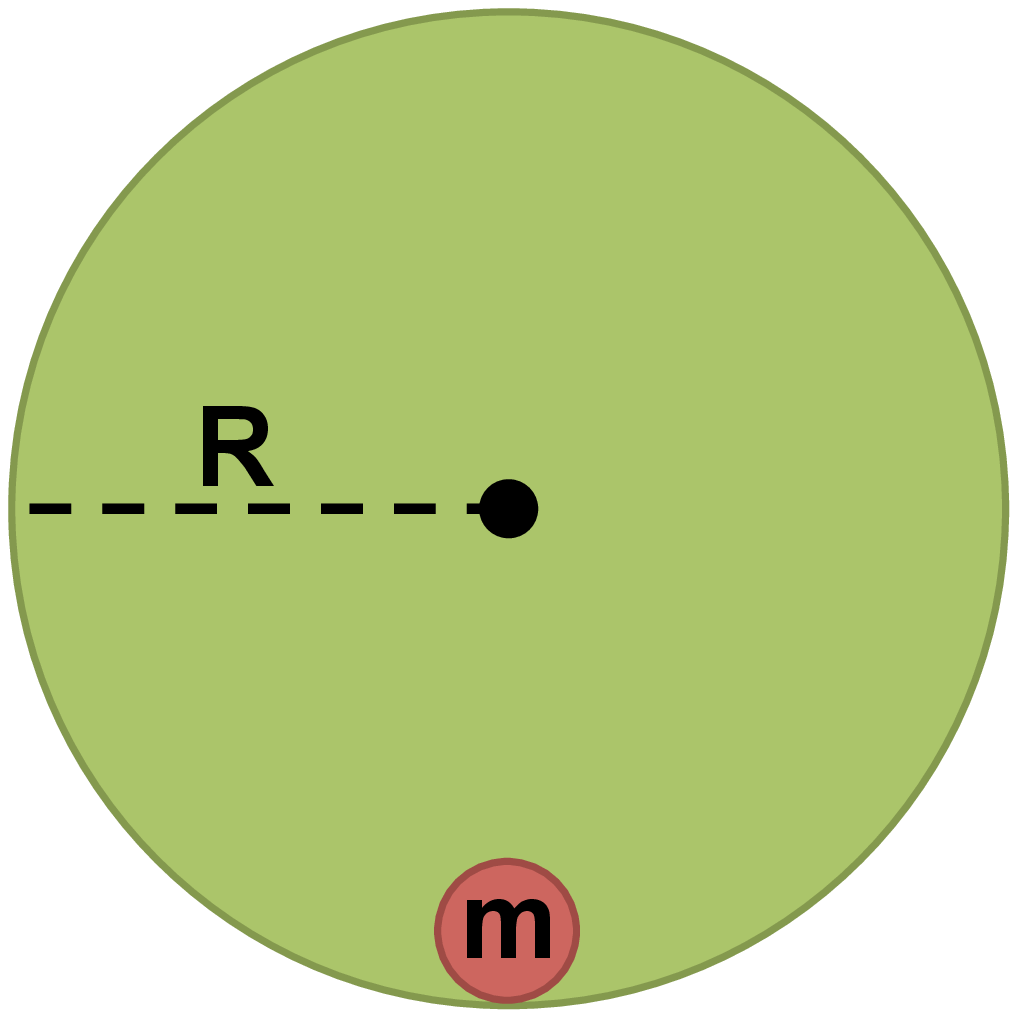}
 				\end{center} 
 				\end{minipage}
				\begin{minipage}{0.60 \linewidth}
				\flushleft (i) Determine $A$ in terms of the given/known constants. What are the units of $A$?\\
				(ii) Find an approximate expression for the gravitational potential energy for $\phi$ near $\phi = 0$. {\it Your expression must be at least 2nd order in $\phi$, and should not contain any trig functions.}
 				\end{minipage}\vspace*{2mm}
        \end{mdframed}
        \end{minipage}
        \label{fig:cylinder}}
        
        \caption{Students' solutions to these Taylor series exam problems were analyzed using the ACER framework. (a) Motion problem developed prior to ACER. (b) Energy problem developed after ACER and used as part of interview studies. Italicized text did not appear on interview documents.}\label{fig:examTS}
\end{figure}

\vspace{-14pt}
\subsubsection{\label{sec:TSmethods}Methods}

Our students are formally exposed to Taylor series expansions (Eqn.\ \ref{eqn:TS}) in mathematics courses taken prior to CM 1. In CM 1, students learn to use Taylor series in problems with physical context. Here, we examine two exam questions that represent typical problems asked of our sophomore students with different contexts: motion and energy. The first problem (Fig. \ref{fig:drag}) was given prior to the development of ACER. It explicitly asks students to perform a Taylor series expansion on an expression for the 1-D position of a particle moving under linear drag. The second exam problem (Fig. \ref{fig:cylinder}) was written after the development of ACER to directly target aspects of Activation and Reflection. Students must find an approximate expression for the gravitational potential energy of a bead sliding inside a frictionless cylinder. 

Exams were collected from two semesters of the course (N=116), each taught by a different instructor. One instructor was traditional research faculty and the other was physics education research faculty involved in the development of transformed course materials. Both instructors made use of these transformed materials. In the first exam study, students (N=45) solved the linear drag problem (Fig.\ \ref{fig:drag}) on the traditional faculty member's first exam. In part i, students were asked to compute the first two terms of a canonical Taylor expansion about $t=0$. Students needed to clearly state the significance of these two terms in part ii. Finally, students needed to perform a Taylor expansion of the same function around $t=m/b$. For the second exam study, students (N=71) were asked to solve the energy problem (Fig.\ \ref{fig:cylinder}) on the PER faculty member's second exam.  In the first part of this problem, students were asked to check the units of the expression for the potential energy. Then, they computed the approximate expression for the potential energy in part ii.

Interview data came from two sets of think-aloud interviews (N=8), performed approximately 1 semester apart. Both studies asked students to solve a number of Taylor series problems, which included formal math and physics questions (e.g., Fig.\ \ref{fig:examTS}). The first study was performed prior to the development of ACER and asked formal math questions first. Physics questions, which were asked at the end of the interview, explicitly cued students to use a Taylor series (e.g., ``perform a Taylor expansion'') and included parts i and ii of the drag question (Fig.\ \ref{fig:drag}). After the development of ACER, it was clear that the first study limited the possibility of observing attempts to process implicit cues. In the second study, formal math problems were moved to the end of the interview and the physics questions contained only implicit cueing (e.g., ``find an approximate expression''). Part ii of the energy question in Fig.\ \ref{fig:cylinder} appeared as part of this study.

\subsubsection{\label{sec:Results2}Results}

This section presents the analysis of student work and the identification student difficulties with Taylor series organized by component and element of the operationalized ACER framework (See Sec.\ \ref{sec:Taylor}).

\textbf{Activation of the tool}: TA1-TA3 are cues embedded in the problem statement that can lead a student to activate resources associated with Taylor expansions, and in some sense, they are organized by the likelihood that they will do so.  The first exam study and think-aloud interview study (Fig.\ \ref{fig:drag}) targeted students' responses to explicit cueing.  Almost all students in the exam study (93\% of 45) attempted a Taylor series on part i of the problem, and most students (87\% of 45) did so again on part iii. Those students who did not attempt a Taylor expansion used some inappropriate form of the binomial expansion (e.g., $(a+b)^n$ rather than $(1+\epsilon)^n$) or skipped part iii. We saw similar success in the first interview study, where no student failed to start the problem with a Taylor expansion when explicitly prompted.  

From the point of view of ACER, the first exam and interview studies limited investigations of Activation to element TA1. The second exam study (Fig.\ \ref{fig:cylinder}), was initially designed to target element TA2 by asking students to ``[f]ind an approximate expression'' in part ii. However, the instructor felt this cueing was too vague, so additional wording was added to the problem statement (the italicized text in Fig.\ \ref{fig:cylinder}). In this study, most students (87\% of 71) attempted a Taylor expansion. Those who did not typically misconstrued the problem by constructing some sort of differential equation (6\% of 71) or left the problem unanswered (6\% of 71). This indicates that students have little trouble activating Taylor series when cued explicitly or implicitly; however, we suspect that the addition of the italicized text in Fig.\ \ref{fig:cylinder} made the cueing more explicit than originally intended.

The second interview study offered a clearer view of students' responses to implicit cueing (TA2) with a question nearly identical to the problem in Fig.\ \ref{fig:cylinder}, but without the italicized text. Two of the four interviewees immediately plugged in the given value (i.e., $\phi$ = 0 in Fig.\ \ref{fig:cylinder}) to determine the approximate expression (e.g., $U(\phi) \approx 0$). Later in the interview, after working through the formal math problems, both participants asked to return to the physics problems and solved them again using Taylor approximations. This suggests that these formal math problems primed the student's resources associated with Taylor series expansions allowing them to connect these resources back to the physics.  A recent study of students' use of Taylor series approximations in the context of statistical mechanics also indicates that upper-division students have difficutly knowing when to use a Taylor expansion when not explicitly prompted to do so \cite{Smith2013}.

\textbf{Construction of the model}: In both the exam and interview studies, the mathematical representation of the physical model was constructed for the students (i.e., $x(t)$ and $U(\phi)$ in Fig.\ \ref{fig:examTS}). However, to compute the Taylor expansion of each function, the physical quantities in each equation had to be mapped onto the general expression for Taylor series (Eqn. \ref{eqn:TS}). Identifying elements TC1--TC4 in a students' written solution was a challenge because students rarely documented their thought process while performing this mapping. When coding for these elements, we focused on how students treated the symbols appearing in each problem. The analysis was holistic, taking into account the full solution that students provided. From this view, nearly all of the exams (study 1, part i - 93\% of 45, N=42; study 1, part iii - 89\% of 45, N=40; study 2 - 89\% of 71, N=63) contained elements from the construction component (i.e., the student did more than superficially manipulate the expressions).

In all studies, every student who attempted a Taylor expansion identified the appropriate symbol as the expansion variable (TC1). Moreover, when determining the expansion point (TC2), most students in the first (93\% of 42) and second (97\% of 63) exam studies had no trouble when this point was zero (i.e., a MacLaurin series). Students often demonstrated their identification of the variable and expansion point through mathematical manipulations (e.g., taking derivatives and constructing functions) or their use of canonical symbolic forms (e.g., $x(t) \approx a + b\,t + c\,t^2$) \cite{sherin2001students}. 
 
\begin{table}
\caption{Difficulties constructing an expansion around a non-zero expansion point -- part iii in Fig.\ \ref{fig:cylinder}. Percentages are of the students who had difficulty with the non-zero expansion point (65\% of 40, N=26). Codes are not exhaustive but represent the most common themes, thus the total N in the table need not sum to 26.}\label{tab:nonzero}
  \begin{ruledtabular}
    \begin{tabular} {l c c}
	\textbf{Difficulty} 								& \textbf{N} 	& \textbf{Percent} 		\\
 	\hline
	Used answer to part i 								& 16 			& 62 					\\
	\hspace{3mm} i.e., $x(t) \approx v_{x0}t+v_{x0}\dfrac{b}{m}t^2$  								& 				& 						\\
	Incorrect functional dependence 					 		& 7 			& 27 			\\
	\hspace{3mm} i.e., $x(t) \approx \dfrac{v_{x0} m}{b} (1-\dfrac{1}{e}) + \dfrac{v_{x0}}{e}\,t - \dfrac{v_{x0}b}{em}\,t^2$ 	& 				& 						\\
    \end{tabular}
  \end{ruledtabular}
\end{table}

While most students correctly identified the expansion around zero, a substantial fraction (47\% of 40, N=19) of students failed to properly identify non-zero expansion points (i.e., $t \approx m/b$ in part iii of Fig.\ \ref{fig:drag}). Most of these students (84\% of 16) simply responded to part iii with their answer for the expansion around $t = 0$ (part i). Students who correctly identified $m/b$ as the expansion point (53\% of 40, N=21) had coefficients in their Taylor expansion that were consistent with evaluating the function and its derivatives at $t=m/b$. Two-thirds of these students (67\% of 21) also had the correct functional dependence (i.e., $(t-m/b)^n$). The remaining one-third (33\% of 21) used the form for an expansion around zero (i.e., $t^n$). Difficulties with constructing an expansion around a non-zero expansion point are summarized in Table \ref{tab:nonzero}.  

Given the specific questions used, our exam studies provided little insight into how students compare the scales of physical quantities (TC3) or how students recast expressions (TC4). All students in the first exam study who attempted a Taylor expansion maintained the already-constructed dimensionless ratio (i.e., $bt/m$) throughout their work. The expression in Fig.\ \ref{fig:drag} was constructed such that the dimensionless ratio appeared in the exponential. In the second study, the expansion variable $\phi$ can be compared to a number directly because it is technically dimensionless. However, follow-up questioning of interviewees provided evidence that students do not have a strong grasp of comparative scales. Only one student in eight clearly articulated that for an expansion to be ``good'', it must be performed over dimensionless variables that are smaller than 1. The other seven students believed their expansion was a ``good'' approximation to the original expressions if the variable (e.g., $t$) was ``small compared to 1'' regardless of the expression under consideration or the presence of a natural comparative scale.  Mathematics education researchers have also observed that some students struggle to identify the range in which an approximation is ``good'', even in purely mathematical problems with no inherent comparative scale \cite{Champney2012}. 

\textbf{Execution of the mathematics}:  Elements TE2--TE4 provide opportunities to capture the type and nature of the mathematical errors made while computing Taylor expansiosn. While the data presented below provide evidence that students made a number of mathematical mistakes, we did not find that such mistakes were the primary barrier to student success.

Almost all students in the exam studies (study 1 - 93\% of 45, N=42; study 2 - 89\% of 71, N=63) performed some mathematical manipulation captured by the execution component (TE1--TE4). Identifying constants and variables (TE1) in the given expressions was only a significant barrier to one student in part i of the first exam study and three students in the second exam study. Students typically demonstrated an awareness of the nature of each symbol by taking derivatives or using an expansion template with the appropriate variable (e.g., $t$ in Fig.\ \ref{fig:drag}). Interview participants often explicitly pointed to symbols and clearly identified them as constants. 

Students computed Taylor expansions through both formal and abbreviated methods (e.g., working through Eqn.\ \ref{eqn:TS} versus using ``expansion templates'' like $\cos x \approx 1 + x^2/2!$). Those who used expansion templates shortcut elements TE2 and TE3 even though these templates stem from taking derivatives of the associated function and evaluating those derivatives around the expansion point. Of students who showed evidence of the execution component, a significant fraction used expansion templates when the expansion was around zero (study 1, part i - 67\% of 42; study 2 - 90\% of 63). The remaining students computed derivatives of the associated functions. 

For part iii of the first exam study, fewer students overall were coded in TE2 and TE3 (57\% of 42, N=24) because a substantial fraction (Table \ref{tab:nonzero}) used their answer for the expansion around $t = 0$ (part i). Of the remaining students, more than two-thirds (71\% of 24) used formal methods to compute their Taylor expansion. This suggests that students are more familiar with templates of MacLaurin expansions. We observed similar trends in our interviews. When confronted with simple functions or expressions, interviewees overwhelmingly elected to use or ask for expansion templates. When simple functions were embedded in more complicated expressions, seven of eight interviewees employed formal methods; only one student used an expansion template. 

The broad ACER framework (Sec.\ \ref{sec:Taylor}) does not capture all the nuances of students' mathematical errors; hence,  we found it constructive to create a number of sub-codes to capture more details. Considering all coded instances of Execution, about one-third (34\% of 147, N=50) contained some mathematical error. Some of these students made only slight algebraic manipulation errors (44\% of 50) such as forgetting  a minus sign or dropping numerical factors. More than half of students with mathematical errors (56\% of 50, N=28) made more serious mistakes, which occurred primarily in part iii of the first exam study. More than half of these students (54\% of 28) made serious expansion mistakes, such as appending variables to ``patch up'' their solutions. That is, students would produce a solution that did not depend on $t$ (e.g., $x(t) = a_0 + a_1\;m/b + ...$) and in the next line append a $t$ (e.g., $x(t) = a_0 + a_1\;m/b\;t + ...$). This was not observed in the interviews, so it is unclear if ``patching up'' an expression represents an error in Construction or Execution, or, possibly, a ``success'' in Reflection. About a quarter (29\% of 28) computed the derivative of the associated functions incorrectly. The remaining students struggled to perform any of the necessary mathematics. Mathematical errors were more prevalent in our interview studies, but few were serious. Of the eight participants, seven made some mathematical mistake, but only one participant computed derivatives incorrectly.

Once the computation is complete, it is typical to organize terms in increasing order (TE4). This practice makes the interpretation of the solution somewhat simpler because terms with similar orders are grouped together and their effect can be discussed together. Most students successfully organized their solution in this way (study 1, part i - 83\% of 42; study 1, part iii - 70\% of 40; study 2 - 97\% of 63). Similarly, all interview participants spontaneously organized their solutions in order of increasing power. However, the practice of organizing solutions did not mean students could readily interpret their solution. As discussed below, many students struggled to make meaningful statements about the physics of their proposed solutions.

\textbf{Reflection on the result}: Once a solution has been constructed, it should be checked for errors and an interpretation should be made. As we discuss below, students rarely offered checks or spontaneously interpreted their solution. When prompted in the second exam study, students checked the units of a solution successfully, but in the first exam study, students struggled to interpret solutions meaningfully.

In the first exam study, no student spontaneously checked their solution to part i or part iii for errors. A check of the units (TR1) would have helped a small fraction of students on part i (10\% of 42), but on part iii, it could have clued more than a third of students (33\% of 40) that something was incorrect about their solution. Part ii of the first exam study (Fig.\ \ref{fig:drag}) forced students to interpret their solution and to connect it to their prior knowledge about motion (TR2). Most students offered little substance in their interpretation. Common responses for the linear term included ``it's the initial v times t'' and ``it's the velocity.'' Only a quarter of students (25\% of 40) mentioned something similar to ``the distance covered in vacuum.'' For the quadratic term, the same fraction of students mentioned that it was the ``drag term" or the ``correction,'' but not a single student mentioned the sign difference between the linear and quadratic terms. In our interviews, no student clearly connected a solution to this problem to the underlying physics.

In the second exam study, students were prompted to check the units of the expression prior to starting the problem (Fig.\ \ref{fig:cylinder}) and most students did this correctly (80\% of 71). Students in the first exam study were not asked to reflect on their solution directly.  Eventually, all four participants in the second think-aloud study produced a solution to the problem that depended on $\phi^2$. They were then asked to discuss any physics that could help them interpret their solution. Only one of the four students made an interpretation of the solution. This student suggested that the system ``looks like a harmonic oscillator,'' and gestured to indicate the oscillation around the bottom of the cylinder. Even with additional prompting by the interviewer, the other three participants expressed only superficial reflections, ``yeah, that looks different [from the original expression].''

\subsubsection{\label{sec:Impl2}Summary \& Implications}

We found that sophomore-level students encountered several challenges when solving Taylor approximation problems. These challenges limited the production of complete, well-articulated solutions. First, knowing when to use Taylor approximations is challenging to students when prompts are less explicit. This difficulty is likely under-represented in our data because we have not explored how students grapple with minimal cueing (TA3). Processing implicit cues is a skill that will follow students throughout their physics careers. Instructors should be aware of what cues they include in problems and how these cues impact student success on Taylor approximations. Second, while students are relatively adept at performing expansions around zero (i.e., MacLaurin series), they struggle to perform Taylor expansions around non-zero expansion points. Difficulties here ranged from failing to demonstrate understanding of expansions around points other than zero to expanding around appropriate points but not producing the correct functional form (Table \ref{tab:nonzero}). Not all Taylor expansions in physics occur around zero, and students must be prepared to solve general expansion problems. Third, sophomore students (like juniors) rarely reflect spontaneously on their solutions. We have commonly observed this challenge for students in all upper-division courses. Checking solutions for errors and constructing meaningful interpretations are practices that are equally important to using mathematics. Yet, these practices are under-emphasized in our current upper-division courses. Problems and activities should be designed to develop students' skills with reflective practices.

\section{\label{sec:Discussion}Summary and Discussion}

We have presented an analytic framework, ACER, that is specifically targeted towards characterizing student difficulties with mathematics in upper-division physics.  The ACER framework provides an organizing structure that focuses on important nodes in students' solutions to complex problems by providing a researcher-guided outline that lays out the key elements of a well-articulated, complete solution.  To account for the complex and highly context-dependent nature of problem solving in advanced undergraduate physics, ACER is designed to be operationalized for specific mathematical tools in different physics contexts rather than as a general description.  We have utilized the operationalized ACER framework to inform and structure investigations of student difficulties with Coulomb's Law and Taylor series. This has allowed us to more clearly identify prevalent difficulties our students demonstrated with each of these topics and to paint a more coherent picture of how these difficulties are interrelated.  

As with any expert-guided description, it should not be assumed a priori that the operationalized ACER framework will span the space of all relevant aspects of actual student problem solving.  It is intended to provide a scaffold from which researchers and instructors who are less familiar with qualitative analysis can ground an analysis of what students actually do when solving mathematically demanding physics problems.  However, additional research comparing the operationalized framework, as produced by the expert task analysis, to interviews and group problem-solving sessions will be necessary to explore the limitations of ACER in terms of capturing emergent aspects of students' work.  

There are several important limitations to the ACER framework.  The framework was designed to target the intersection between mathematics and physics in upper-division physics courses, and it is not well suited to describing student reasoning around purely conceptual or open-ended problems.  Additionally, the framework inherently incorporates some aspects of representation because the translation between verbal, mathematical, graphical, and/or pictorial representations is almost always required to solve physics problems; however, the exact placement of multiple representations within the framework is likely to be highly content dependent.  
Furthermore, we have not commented on the integration of prediction and metacognition into the framework, in part because we rarely observe our students showing explicit signs of either without prompting.  Application of ACER to additional topics and tools will clarify how the framework can shed light on these aspects of problem solving.  

Ongoing projects with ACER include its use to frame investigations of upper-division students' difficulties with delta functions and complex exponentials.  Future work will include analysis of students' difficulties with separation of variables in the context of Laplace's equation.  Each of these projects will facilitate further validation and refinement of ACER as a tool for understanding student difficulties. Future work will also involve leveraging ACER to investigate the evolution of students' difficulties with specific mathematical tools over time.  Specifically, Newton's law of gravity for extended bodies is mathematically very similar to the use of Coulomb's Law for continuous charge distributions but is typically encountered in sophomore physics.  By identifying students' difficulties with gravitation in sophomore classical mechanics and comparing them to difficulties with direct integration in junior electrostatics we will be able to investigate how these difficulties change (or not) as students advance through the curriculum.  

The ACER framework was designed to be a tool not only for researchers but instructors as well.  We have already discussed a number of suggestions for instructors that may help students avoid or overcome the difficulties we identified.  However, ACER can also be used to critique and design problems.  Examining the prompt of a question can identify which components of the framework the problem targets and which ones it might short circuit (e.g., bypassing Activation by instructing the student to use a Taylor series to approximate a function).  This can help instructors to produce homework sets and exams that offer a balanced and complete assessment of all aspects of students' problem solving.

\begin{acknowledgments}
The authors gratefully acknowledge the generous contributions of CU faculty members: A Becker, M. Dubson, E. Kinney, A. Marino, and T. Schibli. This work was funded by NSF-CCLI Grant DUE-1023028, the Science Education Initiative, and a National Science Foundation Graduate Research Fellowship under Award No. DGE 1144083.
\end{acknowledgments}


\bibliography{ACER-refs}
\bibliographystyle{apsper}
\end{document}